# Implementing ECC on Data Link Layer of the OSI Reference Model

Donald Somiari Ene[1], Isobo Nelson Davies[2], Godwin Fred Lenu[3], Ibiere Boma Cookey[4]

[1]Technologist, Information and Communication Technology Centre, Rivers State University, Port Harcourt, Nigeria  
[2]Researcher, Department of Computer Science, Rivers State University, Port Harcourt, Nigeria  
[3]Technologist, Information and Communication Technology Centre Rivers State Univeristy, Port Harcourt, Nigeria  
[4]Lecturer, Department of Computer Science, Rivers State University, Port Harcourt, Nigeria

***Abstract -*** *The Internet, a rapidly expanding communication infrastructure, poses significant cybersecurity challenges. A few techniques have been developed to provide security in the OSI model's application, presentation, and network layers. Instead of using Media Access Control (MAC), this paper recommends using public key Elliptic Curve Cryptography to serve the Data Link Layer. On a microFourQ-MSP-IAR Embedded Workbench IDE-MSP430 7.12.4, a prototype of the architecture was implemented. When the cost of key generation was compared to the curve, it was shown that they are directly related. The cost of key generation time rises in proportion to the bit length of the curve. Because of the large number of connected devices that may be exposed due to a lack of effective security in the data link layer, significant concerns arise in terms of network privacy, governance, and security. As a result, this research suggests using a public key cryptosystem based on an elliptic curve to secure the data link layer for safe data communication in Internet-connected devices.*

## I. INTRODUCTION

Cybersecurity is a continuous effort to safeguard Internet-connected systems and the data they contain from illegal access or harm. It's all about keeping your information private, ensuring that no one other than the intended recipient may view your data when you send it, and keeping our lives and businesses safe [1].

## II. THE OPEN SYSTEMS INTERCONNECTION (OSI)

The Open Systems Interconnection reference model, which is a layered framework conceptualizing how communications should be done between heterogeneous systems, was proposed and developed by the International Standards Organization (ISO) in 1984 to serve as the most basic element of computer networking. The OSI model is an architecture that divides the responsibilities required for system-to-system communication logically.

Each of the OSI model's seven levels has its own level of abstraction, and these layers each fulfill a specific purpose in the communication process [2]. To create the seven levels, the following principles were used.

1) Each layer should perform a well-defined function.
2) A layer should be constructed if a different level of abstraction is required.
3) Layer boundaries should be chosen to reduce information flow across interfaces.
4) Each layer's role should be chosen with the goal of developing internationally recognized protocols.
5) The number of layers should be large enough that distinct functions need to be thrown together in the same layer out of necessary, and small enough that the architecture does not become unwieldy

The layered technique was chosen to break down networking functions into smaller logical chunks. The divide-and-conquer strategy can then be used to isolate and fix the data communication problem [2].

The data link layer is the second of the Open System Interconnection's seven layers, and it is responsible for the reliable passage of data frames from one node to another via the physical layer. This layer is largely responsible for sending raw data to the network layer. Data is communicated by breaking it into little parts known as packets, which are then served to the network layer by the data link layer [3].

The frames are simply the physical hardware address of each network interface card connected to the network. Token Ring, ARCnet, and Ethernet are types of local network data link protocols. The network uses other data link protocols such as Serial Line Internet Protocol (SLIP) or Point-to-Point Protocol (PPP) if data communication extends beyond the local network onto the Internet.

The data connection layer sends blocks of data with the appropriate bit error detection and correction, synchronization, flow control, and error control. Because the physical layer just accepts and transmits a stream of data with no concern for the meaning or structure, this layer is also responsible for constructing and identifying frame boundaries [4].

Due to the fast rise of wireless communications, network security has become a major concern. The Open System International (OSI) model has certain ways for ensuring security at the application, transport, and network layers. Several organizations have banded together to improve security at higher OSI layers, from the presentation layer to the network layer. In any event, the hardening of the Data Link layer is an area that has mostly gone unnoticed [4][5].

The data link layer is one of the layers in the OSI reference model that deals with raw data transmission to the network layer. Data is split into little packets and transmitted over the network. The network layer, which is the layer above it, receives service from the data link layer. The communication channel that connects the connecting nodes is referred to as links, and each datagram must be transferred via a separate link to get from source to destination. The data link layer protocols define the format of packets sent between nodes, as well as functions like error detection, retransmission, and flow control.

The Logical Link Control (LLC) sublayer and the Media Access Control (MAC) sublayer are the two sublayers of the data link layer. According to the Institute of Electrical Electronics Engineers working group 802 (IEEE-802) Local Area Network (LAN) specification, the LLC sublayer's mission is to govern data flow between various services and applications, as well as offer error reporting and acknowledgement mechanisms. The logical link control sublayer subsequently communicates with the MAC sublayers, which are responsible for transporting data across physical media. The LLC sublayer is also responsible for physically addressing the frames. The common types of the MAC sublayer include wired and wireless specifications [5][6].

The three most important functions of the data link layer include;

1) Handling issues that may occur as a result of bit transmission error
2) It controls data flow and set it at a pace that would not overwhelm the sending and receiving devices
3) It helps the transmission of data to the network layer, which is layer 3, where data are logically addressed and routed.

The data link layer, as well as other layers of the OSI model, has a set of defined rules and conventions that are followed for communication to occur. These defined rules and conventions are known as protocols.

The Address Resolution Protocol (ARP) is a protocol that binds an Internet Protocol (IP) address to a network-recognizable physical node address. An ARP request is broadcast by a node when it needs to find a physical MAC address for an IP address. The second device on the network with the IP address responds with its physical address in an ARP message (MAC address). The MAC address table is maintained by each device on the network. The IP address and MAC address of other nodes on the network are stored in this table [7].

Since ARP is a stateless protocol, it will accept an ARP response or reply from another device even if it has not sent an ARP request. This procedure can result in an attacker inserting a bogus entry into a target device's MAC address table, which is known as ARP spoofing or ARP poisoning [7].

This attack vector allows an attacker to impersonate a genuine network node in order to intercept, change, or block data frames being sent over the network. Most of the time, this attack vector is utilized to construct other attack vectors like man-in-the-middle, denial of service (DoS), or session hijacking.

Other attack routes include MAC cloning and Layer 2-based broadcasting, in addition to the ones stated above. The attacker modifies the MAC address table of devices in the network with a phony MAC address in Layer 2-based attack paths. Despite the fact that each network interface cards MAC or physical address is unique globally, an attacker who uses MAC cloner can change it. The attacker accomplishes this by using a DoS attack to disable the target host and then using the target host's layer 2 and layer 3 addresses. These attacks are frequently conducted over a wireless network. The wireless network's signals are broadcast over the air and can be intercepted by an attacker if they are within range. This makes securing the wireless network more complicated and time-consuming [8].

### III. SIGNIFICANCE OF THE RESEARCH

Communication systems have evolved into a need that we use on a daily basis in both our professional and personal lives. It also has an impact on political structures, as we have seen lately. Because this technology is so important to our contemporary way of life, it is critical to analyze the problems that it poses. The Internet will be massively expanded to bridge the digital gap and to enable for the Internet of Things, which will connect tens of billions of devices to the Internet. Only a smart, fast, and secure communication model can succeed.

### IV. IMPLEMENTATION

This section expresses the synthesis results for the system implementation and shows comparison results of a secure data communication in the data link layer in the internet of things devices using Elliptic Curve Cryptography. The purpose for this paper is to describe the activity and procedures adopted to bring the IoT secure communication in the data link layer to bare. The following section shows the synthesis results for the developed system. The tasks involved in this chapter will include identifying hardware and software requirement considering both the functional and non-functional requirements earlier specified.

### V. MATERIALS AND METHODS

All the documented elliptic curve cryptography hardware architectures are connected to the field and elliptic arithmetic necessary, and none of them appear to implement an ECC procedure like digital signature. Although software implementations of elliptic curve techniques have been published, arithmetic operations are slower than hardware solutions and are therefore unsuitable if the cryptosystem needs to conduct a large number of cryptographic operations, such as in a secure web server [9].

LiDIA and MIRACL are two libraries that provide primitives for EC cryptography (Shamus, 2005). Both are C++ libraries that provide a large number of cryptographic primitives and are efficient, general-purpose number-theoretic libraries. Although implementations of EC cryptographic protocols (e.g., ECDSA) and EC generation methods (CM method) based on these libraries appear to exist, they are either not publicly available (LiDIA) or are only partially available (CM method) (MIRACL) [10].

Fundamentally, the efficiency of this system would be dependent on the requirement needed for the expectations to be met. The results are obtained from the ECC application, modified during the research. Most of the results obtained are based on simulation on the MacBook Pro 2019, 16 GB RAM, macOS Big Sur, results may vary if application is simulated on different platform and operating system.

### VI. SETUP AND CONFIGURATION

The IoT Security System core was put together with the specified hardware and software. The model that was used to achieve the result is the Software Reuse Model. The software reuse model is based on C programming, from where some high-performance crypto libraries based on the elliptic curve cryptosystem were used. The parameters for encryption in the architectures shown in the previous section were chosen based on software implementation findings. The characteristics for the arithmetic units, such as finite field and elliptic curve, were obtained from the National Institute of Standards and Technology's recommendations (NIST). The following tools were used to simulate the proposed system:

IAR Embedded Workbench: This is a toolchain that makes a complete IDE integrating all the tools in a single view, guaranteeing reliability, quality, and efficiency in the embedded application. It is a completely integrated development environment incorporating a compiler, assembler, a linker, and a debugger

Simulator: The simulator helped provide a realistic imitation of the operations and controls of the system. This tool simulated the programmable circuit board (PCB) and microcontrollers nicely but as a software.

Library Tools: All the required ISO/ANSI C and C++ source and libraries are included in the ECC libraries. This makes things very easy as there was not need recreating libraries.

ECC Parameters: 192 bits key length as recommended by the National Institute of Standards and Technology (NIST).

The following are the parameters of the encryption simulation.

$a = -3$

$b = 24551555460089438177402939151974517847691080581\\61191238065;$

$p = 62771017353866807638357894232076664160839087003\\90324961279;$

$nB = 2818646689284967968603885680739626753757717668\\7436853369;$

$G = \{6020462823756886567582134805875261119166987663684684818\};$

$Pb = \{28030007865416173313773848974350954991247488818\\90727495642\}$

### VII. ENCRYPTION PROCESS

1) Input text = Who has 1.1.0.1? Tell 1.1.0.178
2) Its equivalent ASCII values are: 87 104 111 32 104 97 115 32 49 46 49 46 48 46 49 63 32 84 101 108 108 32 49 46 49 46 48 46 49 55 56
3) Group the ASCII values with size calculated as Length [IntegerDigits [p, 65536]] – 1, which is gotten as 11. {{87 104 111 32 104 97 115 32 49 46 49}, {46 48 46 49 63 32 84 101 108 108 32}, {49 46 49 46 48 46 49 55 56 32 32}}

4) Convert each group into big integers using FromDigits function with base 65536. {13177360672288210 9440, 31771788121403817984, 1733618950330122240}
5) Pad with 32 at the end of the above list if the number if term is odd, so that pairing can be done. Pair them up as 'Pm', which is one of the parameters used in ECC operation. {13177360672288210 9440, 31771788121403817984, 1733618950330122240, 32}
6) After calculating cipher text, Pc = {kG, Pm + kPb}is obtained as kG={482266228269956459579432612776695653104 9767051156704845687360309 18434}Pm + kPb = {{22453300630458873734750398864264751207 88052604195086791 4569161856562986},
{22453300630458873734750398864264751207 88052604195076791 2750560378271530},
{22453300630458873734750398864264751207 88052604195073787 4581389304575786},
{22453300630458873734750398864264751207 88052604195073614 0962438974453578}}
7) Send the ciphertext Pc to the communicating party. Due to the random k in the encryption operation, every run of the program will create a different ciphertext even with the same input.

## VIII. RESULTS AND DISCUSSION

This section presents execution timing results for the cryptographic scheme ECDSA on the data link layer. The results were obtained simulating the overall system. The architecture was tested using files of the Wireshark logs as data link layer data sets. Signatures where generated and verified; also, files were encrypted and decrypted. The results were compared with results of the radio based wireless system.

### IX. TABLE I

### TIMING RESULTS

| Curve | Key Generation Time (s) | Encryption Time (s) | Decryption Time (s) |
|---|---|---|---|
| NIST 163 | 0.04336 | 0.089972 | 0.087804 |
| NIST 233 | 0.114904 | 0.201624 | 0.174524 |
| NIST 283 | 0.114904 | 0.321948 | 0.321948 |
| NIST 409 | 0.246068 | 0.68292 | 0.592948 |
| NIST 571 | 0.519236 | 1.606488 | 1.660688 |

## X. RESULT INTERPRETATION

Analysis of the cost of encryption against different lengths of curves concludes that they are directly proportional. The length of bit also increases as the increase in the encryption cost; however, this increase is more so rapidly than the cost of key generation. Analysis of the cost of decryption against curves also gives an idea that they are directly proportional. Decryption cost also increases as the increase in bit length of curve; however, this increase is too much exponential as compared with other two costs. It can be concluded that key generation is the least costly process in ECC. Although encryption and decryption take more time, it still takes less time to complete the chain of processes compared to other public key cryptosystem such as RSA, and therefore good for the purpose of securing data link layer communication of the IoT devices. The result so far discussed are also presented graphically in Fig.1 for easy visual representation of the result of encryption cost of NIST curve and time.

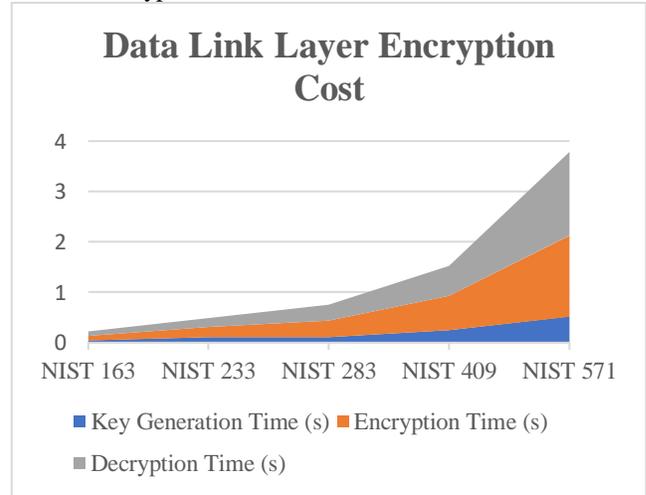

**Fig. 1 Data Link Layer Encryption Cost**

## XI. CONCLUTION

In a world where everything is connected, everything is vulnerable. Data communication and information security in the data link layer of the IoT system has become a serious issue to deal with. Different communication technologies have been continuously integrated with the development of IoT. This includes the radio frequency identification (RFID) systems, wireless sensor network (WSN), mobile vehicle network, etc. However, the communication network corresponding with the environment has become more complex, and the security issues involved are more complex than the existing network infrastructure. Due to the vast number of connected devices that are potentially vulnerable due to the lack of proper security in the data link layer, substantial risks emerge around the issue of privacy, governance, and security in IoT. Therefore, this paper has proposed the use of public key cryptosystem based on elliptic curve to secure the data link layer for secure data communication in IoT devices.